\documentclass[varenna]{cimento}

\usepackage[T1]{fontenc}
\usepackage{amsmath,amsfonts,amssymb,amsthm}
\usepackage{graphicx}

\DeclareMathOperator{\Ramp}{Ramp}
\DeclareMathOperator{\var}{var}
\title{A Dynamical Model for Operational Risk in Banks}
\author{M.~Bardoscia}
\institute{Dipartimento di Fisica, Universit\`a degli Studi di Bari, via Amendola 173, I-70126 Bari, Italy \\
Istituto Nazionale di Fisica Nucleare, Sezione di Bari, via Amendola 173, I-70126 Bari, Italy}

\PACSes{
\PACSit{89.65.Gh}{Economics; econophysics, financial markets, business and management}
\PACSit{02.50.-r}{Probability theory, stochastic processes, and statistics}
}

\begin{document}

\maketitle

\begin{abstract}
Operational risk is the risk relative to monetary losses caused by failures of bank internal processes due to heterogeneous causes. A dynamical model including both spontaneous generation of losses and generation via interactions between different processes is presented; the efforts made by the bank to avoid the occurrence of losses is also taken into account. Under certain hypotheses, the model can be exactly solved and, in principle, the solution can be exploited to estimate most of the model parameters from real data. The forecasting power of the model is also investigated and proved to be surprisingly remarkable.
\end{abstract}

\section{Introduction} \label{sec:intro}
There have been many successful attempts to extend some crucial ideas and techniques of statistical mechanics to plenty of different fields, including several economic topics. The most studied subject in this context is the financial risk, related to the fluctuations of the prices of stocks and other products (see ref.~\cite{mantegna-stanley}), while only more recently new kinds of risks \cite{mcneil-frey-embrechts} like credit risk and operational risk \cite{cruz} have been investigated. In particular the rise of interest in operational risk has started after that the New Basel Capital Accord ~\cite{basel}, also known as Basel II, has prescribed banks to cope with it. 

Operational risk is defined by Basel II as ``the risk of [money] loss [in banks] resulting from inadequate or failed internal processes, people and systems or from external events'' \cite{basel}. In this context the main goal is to determine the \emph{capital requirement}, i.~e. the capital that a bank has to put aside every year to cover operational losses. The capital requirement is usually identified with the Value-at-Risk (VaR) over the time horizon of one year with level of confidence $99.9\%$, defined as the $99.9$ percentile of the yearly loss distribution, meaning that a loss larger than the VaR occurs with probability $0.01$ in one year.

Perhaps the most widespread approach to operational risk is the Loss Distribution Approach (LDA) \cite{frachot-moudoulaud-roncalli}. In the context of the LDA losses are classified by the business line in which the loss occurs and by its cause in $56$ couples; the loss distribution of each couple is fitted from data of historical losses assuming that no correlations exist between the losses occurred in different couples. However it is easy to provide an example to show that such an hypothesis is not realistic; let us suppose that a failure occurs in the transaction control system at the time $t_1$ and repaired at the time $t_2 > t_1$, generating a loss equal to the cost of reparation; however from the time $t_1$ to the time $t_2$ some transactions may fail or may be wrongly authorized, resulting in other losses. The example shows that a crucial mechanism for the generation of losses is given by interactions which are non local in time (the time interval $[t_1, t_2]$ may last months) and non symmetrical (a failed transaction does not cause a failure in the transaction control system). There are several proposal to include the correlations among different couples in the LDA (see refs.~\cite{aue-kalkbrener, boecker-klueppelberg, gourier-farkas-abbate}), but no one has reached a general consensus. Moreover the LDA is limited to give a static and purely statistical description of the losses, independent from the dynamical mechanisms behind their generation.

The approach presented in this contribution is based on a totally different framework \cite{leippold-vanini}: the bank is regarded as a dynamical system whose degrees of freedom are variables representing the losses occurred in different processes (that can be thought as abstractions of the couples of the LDA) whose state is updated according to an equation of motion that includes several mechanisms for the generation of losses.

\enlargethispage{\baselineskip}

\section{The Model}
The model can be considered a generalization of the one introduced in refs.~\cite{kuhn-neu, anand-kuhn} and consists of $N$ positive real variables $l_i(t)$, $i = 1, \dots, N$, representing the amount of the monetary loss occurred at the time $t$ in the $i$-th process. The reason for defining $l_i(t)$ positive is that the databases of operational losses collected by banks have only positive entries: in other words the observable quantity is intrinsically positive. In the context of operational risk the most important quantity is the cumulative loss up to the time $t$: $z_i(t) = \sum_{s \leq t} l_i(s)$, since it can be taken as a measure of the capital requirement over the time horizon $t$. The values of the variables $l_i(t)$ are updated according to a discrete time equation of motion that includes two different mechanisms for the generation of losses: spontaneous generation via a noise term and interaction with other processes; the possibility that the banks invest a fixed amount of money for unit of time to keep a process working is also taken into account. The equation of motion is:
\begin{equation} 	\label{eq:motion}
	l_i(t) = \Ramp \left[ \sum_{j=1}^N J_{ij} \sum_{s=1}^{t_{ij}^*} \Theta \left[ l_j(t-s) \right] \; + \; \theta_i \; + \; \xi_i(t) \right] ,
\end{equation}
where $\Ramp(x) = x$ for $x > 0$ and it is equal to zero elsewhere, while $\Theta(x) = 1$ for $x > 0$ and it is equal to zero elsewhere; the ramp function ensures that $l_i(t)$ stays positive at all the times $t$. As it can be seen from eq.~(\ref{eq:motion}), the value of $l_i(t)$ depends on the interplay among the terms of the argument of the ramp function: the positive terms tend to generate a loss, while the negative terms tend to avoid the occurrence of a loss.

The first term accounts for \emph{potential} losses generated from the interaction with other processes and it is build in the following way: if $J_{ij} > 0$, each loss occurred between the time steps $t - t_{ij}^*$ and $t - 1$ in the $j$-th process generates a \emph{potential} loss of amount $J_{ij}$ in the $i$-th process at time $t$; $t_{ij}^*$ measures how much non-local in time the coupling between the $i$-th and the $j$-th process is; in general both $J_{ij}$ and $t_{ij}^*$ are not symmetrical. The noise term $\xi_i(t)$ accounts for the spontaneous generation of losses (like those generated by failures or human errors) and thus must have a positive support; it is $\delta$-correlated in time, does not depend on time and its distribution is exponential: $\rho( \xi_i ) = \lambda_i e^{-\lambda_i \xi_i}$; provided that the variance of $\xi_i(t)$ is finite, the qualitative results do not depend on its distribution and the quantitative results in ref.~\cite{bardoscia-bellotti} can be easily extended. If $\theta_i < 0$, it can be interpreted as the amount of money per unit of time invested on the $i$-th process to keep it running: in fact the sum of the the interaction term and the noise has to be greater than a threshold equal to $|\theta_i|$ to effectively generate a loss.

In ref.~\cite{bardoscia-bellotti} it is shown that the model can be exactly solved, in the sense that all the moments of the distribution of $l_i(t)$ can be calculated, provided that the matrix of couplings $J$ satisfies the following hypothesis. Let us associate to each process a node in a graph and, if $J_{ij} \neq 0$, i.~e. if the state of the $i$-th process is influenced by the state of the $j$-th process, let us draw a directed edge starting from the $j$-th node and ending to $i$-th node; if such graph has no loops, i.~e. if it is a directed acyclic graph (see ref.~\cite{thulasiraman-swamy} for some basic definitions about graphs), then the matrix of couplings $J$ is said to have no casual loops, and all the moments of $l_i(t)$ can be calculated. If $J$ has no causal loops it is also true that  $z_i(t)$ is the sum of independent and identically distributed variables of finite variance and thus, via the central limit theorem, the asymptotic distribution of $z_i(t)$ is Gaussian with $\langle z_i(t) \rangle = t \, \langle l_i(t) \rangle$ and $\var z_i(t) = t \, \var l_i(t)$. This is one of the results still holding for any distribution of $\xi_i(t)$, provided that its variance is finite.

\enlargethispage{\baselineskip}

\section{Parameters Estimation} \label{sec:estimation}
In ref.~\cite{bardoscia-bellotti} it is shown that, in principle, some of the model parameters can be estimated from real data, i.~e. from a database of operational losses. Such a database is a collection of the past losses registered inside the bank and, in order to be suitable for the estimation of the model parameters, for each loss both the process in which it has occurred and the time at which it has occurred must have been recorded. In the estimation procedure the inverse of the frequency with which the losses have been recorded in the database is taken as the length of a time step of the model, so that the database of historical losses can be interpreted as a realization of eq.~\eqref{eq:motion}.

There two possible approaches to the estimation and in both of them the matrix $t^*$ must be known; in the first one $\lambda$ must be also known and $\theta$ and $J$ are estimated; in the second one $J$ must have no causal loops and, in addition to $\theta$ and to the non zero elements of $J$, the exact solution is exploited to estimate also $\lambda$; it has to be stressed that the knowledge of the graph associated with $J$ implies the knowledge only about which elements of $J$ are equal to zero. Let us point out that some constraints on the possible values of the estimated parameters exist: for both the estimation approaches $\theta_i$ must be negative, which is precisely the case we are interested in. Additional bounds on the values of the elements of $J$ exist (see ref.~\cite{bardoscia-bellotti}) and their interpretation is that the control exerted by the bank on the processes via $\theta_i$ is so strong that the interactions alone (without the noise) are not sufficient to generate a loss.

\enlargethispage{\baselineskip}

\section{Forecasting Power and VaR}
The forecasting power of the model is investigated using a simulated database of operational losses; the first step is to generate a trajectory (that will be called original trajectory) of $T$ time steps from eq.~(\ref{eq:motion}), to interpret it as a database of operational losses and to estimate the parameters only from the first $f \, T$ time steps ($0 < f \leq 1$); the second step is to use the estimated parameters to calculate $\langle z_i(t) \rangle$ and $\var z_i(t)$ by means of the exact solution (in the case in which $J$ has no casual loops) or by sampling a great number of trajectories from eq.~(\ref{eq:motion}) and compare it to the original trajectory, also to the part not used to estimate the parameters. If $f = 1$ it reduces to a validation test for the estimation of the parameters. 

\begin{figure}[t]
	\centering
	\includegraphics[width=0.48\columnwidth]{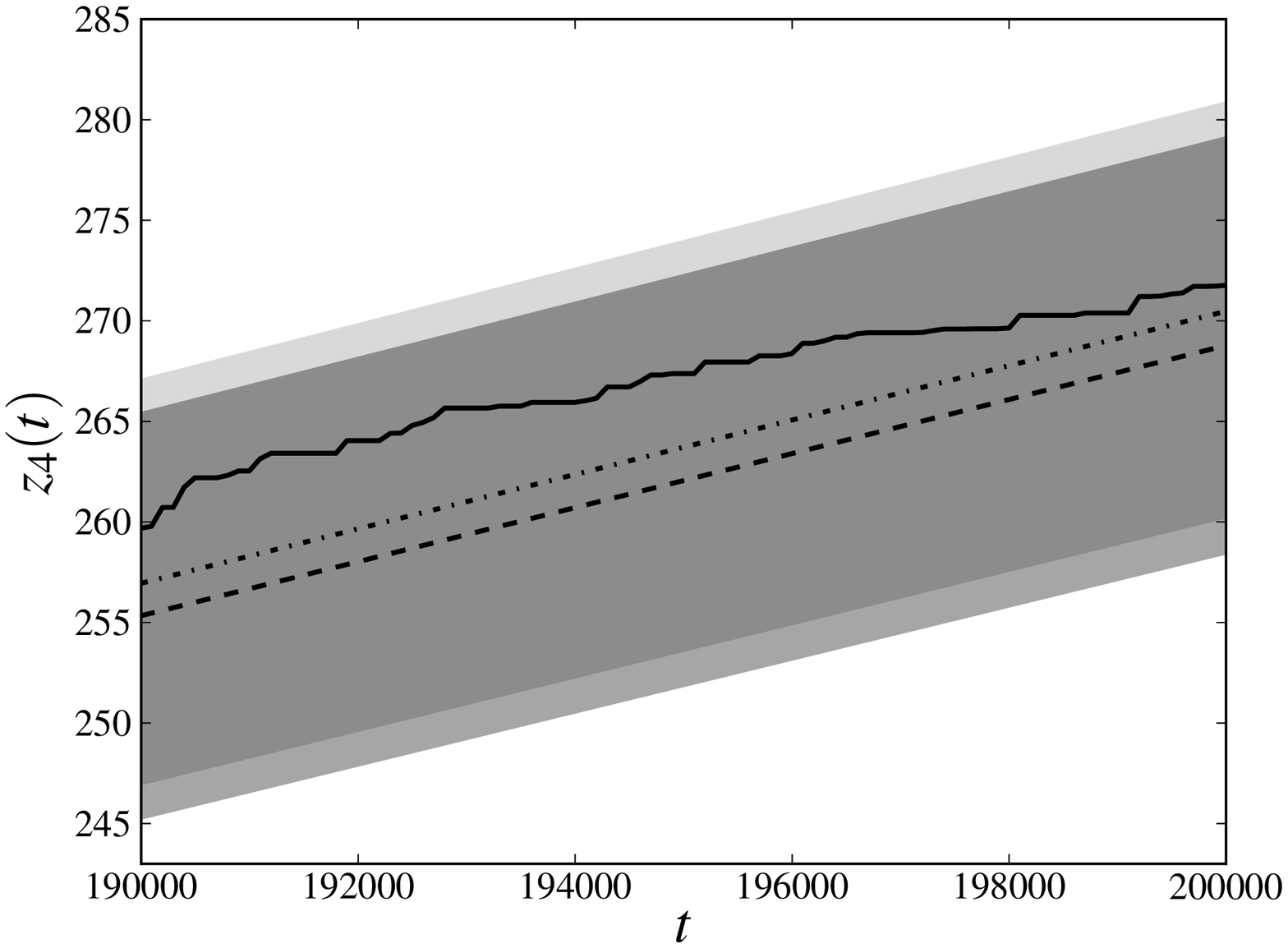}
	\includegraphics[width=0.48\columnwidth]{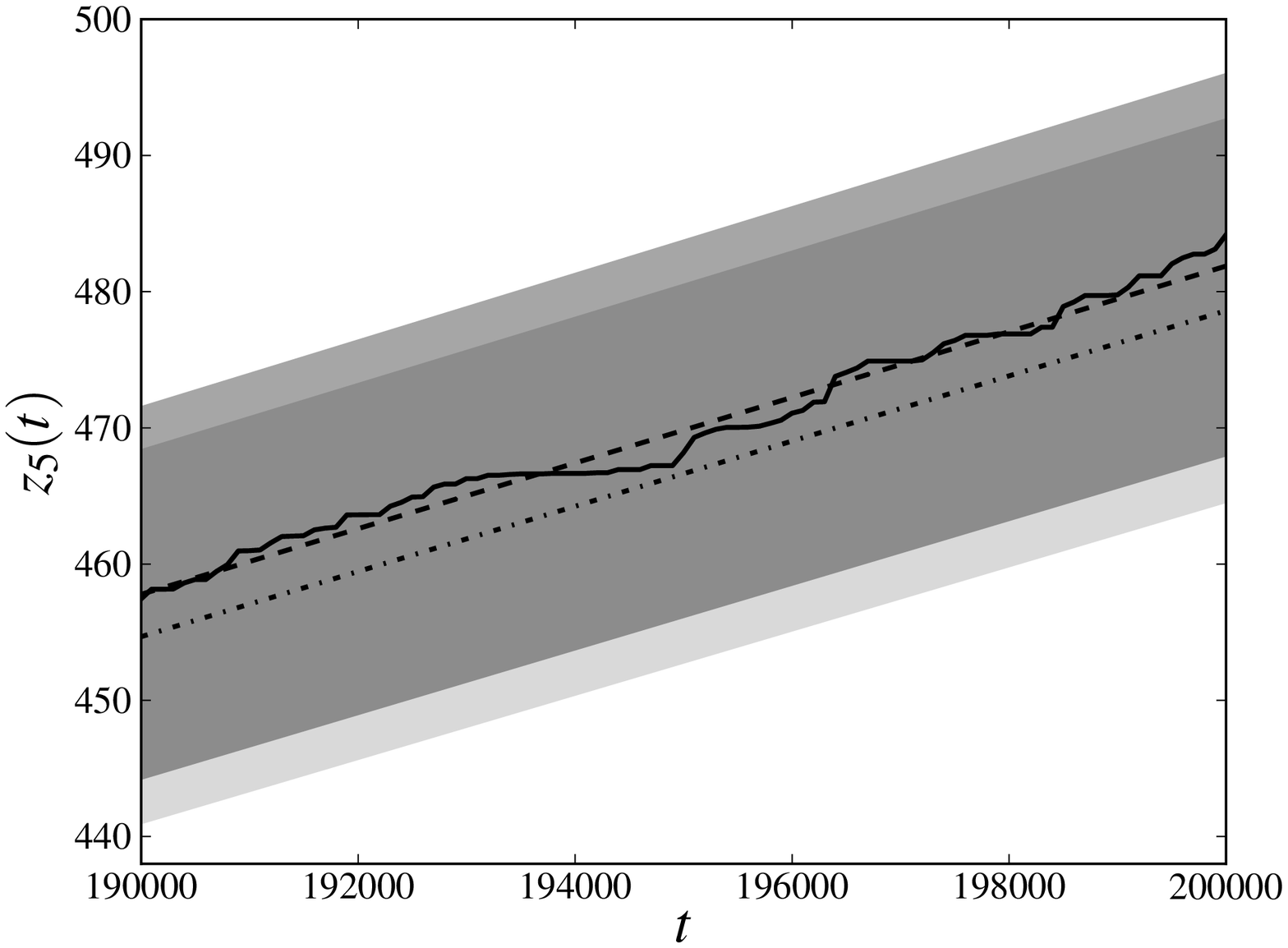}	
	\caption{Cumulative loss of the original trajectory (solid line) and $\langle z_i(t) \rangle$, the average of $z_i(t)$ obtained estimating the parameters from the original trajectory, for $f = 1$ (dashed line) and $f = 0.75$ (dash-dotted line); the limits of the semi-transparent regions are $\langle z_i(t) \rangle \pm \sigma_{z_i}(t)$, for $f = 1$ (dark grey) and $f = 0.75$ (light grey); $z_i^*(t)$ is reproduced with an uncertainty which is far less than $\sigma_{z_i}(t)$ and the error regions overlap almost completely.}
	\label{fig:cumul}
\end{figure}

The whole procedure is carried over in ref.~\cite{bardoscia-bellotti} for $N = 5$, using the first approach to the estimation of parameters: we suggest to consult that reference for all the details, including the parameters used to generate the original trajectory which have been chosen to be compatible with the bounds illustrated in section~\ref{sec:estimation}. Here we present only the results relative to the process $4$ and $5$ discussed in ref.~\cite{bardoscia-bellotti}, the processes with the most complicated interactions. In fig.~\ref{fig:cumul} it is shown that the cumulative loss relative to the original trajectory is indistinguishable from $\langle z_i(t) \rangle$ within an error smaller than $\sigma_{z_i}(t) = \sqrt{\var z_i(t)}$, both for $f = 0.75$ and for $f = 1$, showing that the model has a remarkable forecasting power. Analogous results are obtained for the other processes.

As briefly discussed in section \ref{sec:intro}, the most widely used measure of the capital requirement is the VaR with level of confidence $99.9$ over the time horizon of one year. For our model the VaR over the time horizon $t$ is the $99.9$ percentile of the distribution of $z_i(t)$ and, since the distribution of $z_i(t)$ is Gaussian, for large $t$ the VaR is approximately equal to $\langle z_i(t) \rangle +  3 \, \sigma_{z_i}(t)$; once the link between the length of a time step and the real time has been established as pointed out in section~\ref{sec:estimation}, the VaR over the desired time horizon can be calculated. However in our case the estimation procedure has been carried out with simulated data and, since no real time scale is available, it is reasonable to calculate the VaR over the time horizon $T$. In ref.~\cite{bardoscia-bellotti} it is shown that the relative error between the VaRs over the time horizon $T$ relative to $f = 1$ and $f = 0.75$ is $\simeq 10^{-3}$ for all the processes, showing that also the capital requirement can be reliably forecast.

\enlargethispage{\baselineskip}

\acknowledgments
The author would like to thank R.~Bellotti for the fruitful discussions and M.~V.~Carlucci for the countless suggestions and her precious help.

\end{document}